\documentclass[11pt,reqno]{amsart}

\usepackage{amssymb}
\usepackage[reqno]{amsmath}
\usepackage{amscd}

\newcommand{\pre}[1]{\hspace{1pt}{}^{{#1}}\hspace{-1pt}}

\begin{document}

\begin{flushright}
\baselineskip=12pt
CERN--TH/98--252\quad
\end{flushright}

\title[Matter fields]{Canonical general relativity:\\
matter fields in a general linear frame}
\author{M. A. Clayton}
\address{CERN--Th, CH--1211 Geneva 23, Switzerland}
\email{Michael.A.Clayton@cern.ch}
\date{\today}

\begin{abstract}
Building on the results of previous work~\cite{Clayton:1997c}, we demonstrate how matter fields are incorporated into the general linear frame approach to general relativity.
When considering the Maxwell one-form field, we find that the system that leads naturally to canonical vierbein general relativity has the extrinsic curvature of the Cauchy surface represented by gravitational as well as non-gravitational degrees of freedom.
Nevertheless the metric compatibility conditions are undisturbed, and this apparent derivative-coupling is seen to be an effect of working with (possibly orthonormal) linear frames.
The formalism is adapted to consider a Dirac Fermion, where we find that a milder form of this apparent derivative-coupling appears.
\end{abstract}
\maketitle

\section{Introduction}
\label{sect:Intro}

This work is essentially an extension of that presented in~\cite{Clayton:1997c}, showing that the Hamiltonian description of matter fields minimally coupled to general relativity (GR) is compatible with the general linear frame formalism.
In~\cite{Clayton:1997c} we gave a detailed description of the surface geometry and canonical structure of general relativity in a general linear frame, using two different choices of the diffeomorphism constraints: the `unprimed' constraints which are compatible with a coordinate frame gauge choice, and the `primed' constraints which are compatible with an orthonormal (or Lorentz) frame gauge choice.

Briefly, we began with an action that allowed one to treat the metric and vierbein fields independently.
Then, specializing to a surface-normal frame, we developed a Hamiltonian formalism that allowed one to consider the spatial frame independently from the spatial metric degrees of freedom as initial data, each with its own conjugate momenta.
(Note that we have followed~\cite{Clayton:1996b} and specialized somewhat by choosing a surface-normal frame and assuming the existence of a spatial coordinate frame, choosing a minimal representation of the evolution of the spatial hypersurface in spacetime---for alternative points of view, see~\cite{Charap+Henneaux+Nelson:1988,Nelson+Regge:1986}.
Operationally this should be seen as no particular limitation since any numerical scheme would certainly begin from such a point of view.)
We naturally end up with atlas fields $(N,N^a,{N^a}_b)$ that enforce the Hamiltonian, momentum and frame constraints respectively.
The latter constraints ${\underline{\boldsymbol{\mathcal{J}}}^a}_{b}$ generate infinitesimal frame transformations, and are common to the two choices of diffeomorphism generators considered:
the unprimed  $\underline{\boldsymbol{\mathcal{H}}}$ and $\underline{\boldsymbol{\mathcal{H}}}_a$ and primed $\underline{\boldsymbol{\mathcal{H}}}^\prime$ and ${\underline{\boldsymbol{\mathcal{H}}}^\prime}_a$ constraints, related by nontrivial factors involving ${\underline{\boldsymbol{\mathcal{J}}}^a}_{b}$.

We shall see that, as expected, the addition of matter fields poses no great difficulty.
The somewhat surprising feature that appears is the breakdown of DeWitt's ``Riemannian Structure''~\cite{Kuchar:1977}, as matter fields enter into the definition of the momentum conjugate to the frame fields (this should appear for any tensor field that behaves non-trivially under frame transformations).
This is operationally due to the presence of derivatives of the frame field in the matter action, entering through the non-vanishing structure `constants' ${C_{AB}}^C$ which appear.
Note that even though the gravitational and matter sectors are mixed, we show that the evolution of the spatial metric is identical to that required by the compatibility equations (\textit{i.e.}, the definition of the extrinsic curvature is unaltered) and the field equations are equivalent to those derived in a coordinate frame.
Thus this form of derivative-coupling is benign in the sense that the light cone structure of the theory is undisturbed~\cite{Kuchar:1977}---trivially because it is a coordinate frame theory that has been re-written.
Despite na\"{\i}ve expectation, this works because the matter stress-energy tensor remains independent of second-derivatives of the gravitational variables.
We also find that the Einstein-Dirac (E-D) system is also derivative-coupled in this sense, and may be easily treated via a restriction of the formalism.

The ``unprimed'' system of constraints is closely related to the coordinate frame approaches: evolution and Lie transport act solely on the components of tensors and the frame plays a passive role.
To be specific, infinitesimal spatial diffeomorphisms are defined to leave the frame unaffected and change the tensor components in the usual manner~\cite{Wald:1984} $\xi:T_a\rightarrow\pounds_\xi[T]_a$.
Writing Einstein's equations in terms of the components of the metric $\gamma_{ab}$, frame ${E^i}_a$ and extrinsic curvature $K_{ab}$ of a spacelike hypersurface $\Sigma$, we have
\begin{subequations}\label{eq:unprimed EE}
\begin{gather}
\label{eq:vac perped fields}
\mathrm{d}_\perp [\gamma]_{ab}=2K_{ab},\quad
\mathrm{d}_\perp [{E^i}_a]=0,\\
\mathrm{d}_\perp [K]_{ab}=-R_{ab}
+\nabla_a\nabla_b[N]/N
-KK_{ab}+2K_{ac}{K^c}_{b}
+\tfrac{1}{2}\bar{T}^{\textsc{m}}_{ab},
\end{gather}
\end{subequations}
where $\mathrm{d}_\perp$ is the general linear frame generalization (see the discussion in Section~\textrm{III}-B of~\cite{Clayton:1997c} for more details) of the surface-covariant normal derivative operator~\cite{Isenberg+Nester:1980}.
We have added a matter stress-energy contribution $\bar{T}_{AB}:=T_{AB}-\tfrac{1}{2}\mathrm{g}_{AB}T$ to~\eqref{eq:unprimed EE} for later reference, and we fix $16\pi\mathrm{G}=\mathrm{c}=1$.
Lowercase Greek and Latin letters represent spacetime and spatial coordinate components and upper and lower-case Roman letters indicate spacetime and spatial frame components respectively.
We also use $\pre{4}\nabla_A$ for the spacetime covariant derivative and $\nabla_a$ for the intrinsic surface covariant derivative; more details may be found in~\cite{Clayton:1997c}.

In the Hamiltonian system the action of $\pounds$ is represented on phase space by the momentum constraint $\underline{\boldsymbol{\mathcal{H}}}_a$ as $\pounds_{\vec{\xi}}[\cdot]\rightarrow\{\cdot,\underline{\boldsymbol{\mathcal{H}}}_a[\xi^a]\}$.
We use the notation, for example, $\underline{\boldsymbol{\mathcal{H}}}_a[\xi^a]:=\int_\Sigma d^3x\,\xi^a\underline{\boldsymbol{\mathcal{H}}}_a$, and $\{\cdot,\cdot\}$ are the Poisson brackets given by ($32$) of~\cite{Clayton:1997c}.
The Hamiltonian constraint $\underline{\boldsymbol{\mathcal{H}}}$ represents $\mathrm{d}_\perp$ on phase space, and reproduces~\eqref{eq:unprimed EE} weakly:
\begin{subequations}\label{eq:vacuum unprimed}
\begin{gather}
\label{eq:unp first}
\bigl\{\gamma_{ab},\underline{\boldsymbol{\mathcal{H}}}^{\textsc{gr}}[f]\bigr\}
=2fk_{ab},\quad
\bigl\{{E^i}_a,\underline{\boldsymbol{\mathcal{H}}}^{\textsc{gr}}[f]\bigr\}
=0,\\
\label{eq:unp second}
\bigl\{k_{ab},\underline{\boldsymbol{\mathcal{H}}}^{\textsc{gr}}[f]\bigr\}
=-fR_{ab}
+\nabla_a\nabla_b[f]
-fkk_{ab}+2fk_{ac}{k^c}_{b}
-\tfrac{1}{4}f\gamma_{ab}\mathcal{H}^{\textsc{gr}},\\
\label{eq:unp third}
\bigl\{k^\prime_{ab},\underline{\boldsymbol{\mathcal{H}}}^{\textsc{gr}}[f]\bigr\}
=-fR_{ab}
+\nabla_a\nabla_b[f]
-fkk^\prime_{ab}
+2f{k_{(a}}^ck^\prime_{cb)}
-\tfrac{1}{4}f\gamma_{ab}\mathcal{H}^{\textsc{gr}}
+f{k_{(a}}^c\mathcal{J}^{\textsc{gr}}_{[cb)]}.
\end{gather}
\end{subequations}
(We denote the symmetric and antisymmetric projection on any pair of indices by, for example, $T_{(ab)}:=\tfrac{1}{2}(T_{ab}+T_{ba})$ and $T_{[ab]}:=\tfrac{1}{2}(T_{ab}-T_{ba})$, densities with respect to the spatial metric by boldfaced symbols, and densities with respect to the determinant of the spatial vierbein with an underline.)
We found that the extrinsic curvature $K_{ab}$ was equivalently represented on phase space by the two tensors ($\underline{\pi}_{ab}$ and ${\boldsymbol{p}^a}_i$ are the momenta conjugate to $\boldsymbol{\gamma}^{ab}$ and ${\underline{E}^i}_a$ respectively)
\begin{equation}\label{eq:k defs}
k_{ab}:=-(\pi_{ab}-\tfrac{1}{4}\gamma_{ab}\pi),\quad
k^\prime_{ab}:=-\tfrac{1}{2}\gamma_{(ac}{p^c}_i{E^i}_{b)},
\end{equation}
and the weak equivalence of~\eqref{eq:unp second} and~\eqref{eq:unp third} guarantee that the evolution equations reproduce Einstein's equations~\eqref{eq:unprimed EE}.
Since we are working with a general linear frame on $\Sigma$ we also have the generators of frame transformations ${\underline{\boldsymbol{\mathcal{J}}}^a}_b$, which act, for example, as $\{T_a,{\underline{\boldsymbol{\mathcal{J}}}^b}_c[{\omega^c}_b]\}=\Delta_{\tilde{\omega}}[T]_a=-{\omega^b}_aT_b$, and satisfy the Lie algebra of $\mathfrak{gl}(3,\mathbb{R})$
\begin{equation}\label{eq:Jays}
\bigl\{{\underline{\boldsymbol{\mathcal{J}}}^a}_{b},
{\underline{\boldsymbol{\mathcal{J}}}^c}_{d}[{\omega^d}_{c}]\bigr\}
=\Delta_{\tilde{\omega}}\bigl[{\underline{\boldsymbol{\mathcal{J}}}\bigr]^a}_{b}.
\end{equation}
Note that $\mathcal{J}^{\textsc{gr}}_{(ab)}=-2(k_{ab}-\gamma_{ab}k)+2(k^\prime_{ab}-\gamma_{ab}k^\prime)$, which guarantees that $k_{ab}\approx k^\prime_{ab}$.
The constraint algebra consists of~\eqref{eq:Jays} and
\begin{subequations}\label{eq:vacuum unprimed algebra}
\begin{gather}
\label{eq:vua first}
\bigl\{\underline{\boldsymbol{\mathcal{H}}}[f],
{\underline{\boldsymbol{\mathcal{J}}}^a}_{b}[{\omega^b}_{a}]\bigr\}=0,\quad
\bigl\{\underline{\boldsymbol{\mathcal{H}}}_a[f^a]
,{\underline{\boldsymbol{\mathcal{J}}}^a}_{b}[{\omega^b}_{a}]\bigr\}
=\int_\Sigma d^3x\,
f^a\Delta_{\tilde{\omega}}[\underline{\boldsymbol{\mathcal{H}}}]_a,\\
\label{eq:vua second}
\bigl\{\underline{\boldsymbol{\mathcal{H}}}[f],
\underline{\boldsymbol{\mathcal{H}}}[g]\bigr\}
=(f\nabla_a[g]-g\nabla_a[f])\gamma^{ab}\underline{\boldsymbol{\mathcal{H}}}_b,\\
\label{eq:vua third}
\bigl\{\underline{\boldsymbol{\mathcal{H}}}[f],
\underline{\boldsymbol{\mathcal{H}}}_a[g^a]\bigr\}
=\int_\Sigma dx\, 
\underline{f}\pounds_{\vec{g}}[\underline{\boldsymbol{\mathcal{H}}}],\quad
\bigl\{\underline{\boldsymbol{\mathcal{H}}}_a[f^a],
\underline{\boldsymbol{\mathcal{H}}}_b[g^b]\bigr\}
=\int_\Sigma d^3x\, 
\underline{f}^a\pounds_{\vec{g}}[\underline{\boldsymbol{\mathcal{H}}}]_a, 
\end{gather}
\end{subequations}
which was also derived in~\cite{Clayton:1996b} using the geometric arguments of Teitelboim~\cite{Teitelboim:1973,Hojman+Kuchar+Teitelboim:1976}, and which is a fairly straightforward generalization of that found elsewhere (see for example~\cite{Isenberg+Nester:1980}).

We then noted from the form of~\eqref{eq:vac perped fields} or~\eqref{eq:unp first} that these generators are not very convenient if one wants to consider the restriction to orthonormal frames on $\Sigma$, since if one has chosen $\gamma_{ab}=\delta_{ab}$, the actions of $\mathrm{d}_\perp$ and $\pounds$ (and therefore $\underline{\boldsymbol{\mathcal{H}}}$ and $\underline{\boldsymbol{\mathcal{H}}}_a$) do not leave this form invariant.
There are, however, alternative representations of infinitesimal diffeomorphisms and normal evolution in which the spatial frame plays a more active role.
In particular, we represent an infinitesimal diffeomorphism by $\pounds^\prime$ which acts on a frame as $\pounds^\prime_{\vec{\xi}}[{E^i}_a]=\Delta_{\tilde{\nabla\xi}}[{E^i}_a]$ and on the components of tensors as the covariant derivative $\pounds^\prime_{\vec{\xi}}[T]_a=\xi^b\nabla_b[T]_a$; so that the action on the tensor itself is identical to that of $\pounds$: $\pounds_{\vec{\xi}}[T_a\theta^a]=\pounds^\prime_{\vec{\xi}}[T_a\theta^a]$.
The result is that the action of $\pounds^\prime$ on the components of the spatial metric vanishes due to metric compatibility, and so its action is consistent with the limit to orthonormal spatial frames---note though that it is no longer consistent with the limit to a coordinate frame.
Similarly we define the operator $\mathrm{d}^\prime_\perp$ to act of the frame as $\mathrm{d}^\prime_\perp[{E^i}_a]=\Delta_{\tilde{K}}[{E^i}_a]=-{K^b}_a{E^i}_b$ and on the components of tensors as $\mathrm{d}^\prime_\perp=\mathrm{d}_\perp+\Delta_{\tilde{K}}$.
Using these, Einstein's equations~\eqref{eq:unprimed EE} appear as
\begin{subequations}\label{eq:primed EE}
\begin{gather}
\label{eq:prime evol comp}
\mathrm{d}^\prime_\perp [\gamma]_{ab}=0,\quad
\mathrm{d}^\prime_\perp [{E^i}_a]
=\Delta_{\tilde{K}}{E^i}_b
=-{K^b}_a{E^i}_b,\\
\mathrm{d}^\prime_\perp [K]_{ab}=-R_{ab}
+\nabla_a\nabla_b[N]/N
-KK_{ab}
+\tfrac{1}{2}\bar{T}^{\textsc{m}}_{ab}.
\end{gather}
\end{subequations}

The action of $\mathrm{d}^\prime_\perp$ and $\pounds^\prime$ are represented on phase space by the primed Hamiltonian constraint $\underline{\boldsymbol{\mathcal{H}}}^\prime$ and the primed momentum constraint $\underline{\boldsymbol{\mathcal{H}}}^\prime_a$ respectively (which correspond to a particular choice of mixing of the unprimed constraints with the frame rotation generators), giving
\begin{subequations}\label{eq:vacuum primed}
\begin{gather}
\label{eq:pr first}
\bigl\{\gamma_{ab},\underline{\boldsymbol{\mathcal{H}}}^{\prime\textsc{gr}}[f]\bigr\}
=0,\quad
\bigl\{{E^i}_a,\underline{\boldsymbol{\mathcal{H}}}^{\prime\textsc{gr}}[f]\bigr\}
=\Delta_{\tilde{k}^\prime}[{E^i}_a]=-{{k^\prime}^b}_a{E^i}_b,\\
\label{eq:pr second}
\bigl\{k_{ab},\underline{\boldsymbol{\mathcal{H}}}^{\prime\textsc{gr}}[f]\bigr\}
=-fR_{ab}
+\nabla_a\nabla_b[f]
-fk^\prime k_{ab}
-\tfrac{1}{4}f\gamma_{ab}\mathcal{H}^{\prime\textsc{gr}}
-{{k^\prime}_{(a}}^c\mathcal{J}^{\textsc{gr}}_{[b)c]},\\
\label{eq:pr third}
\bigl\{k^\prime_{ab},\underline{\boldsymbol{\mathcal{H}}}^{\prime\textsc{gr}}[f]\bigr\}
=-fR_{ab}
+\nabla_a\nabla_b[f]
-fk^\prime k^\prime_{ab}
-\tfrac{1}{4}f\gamma_{ab}\mathcal{H}^{\prime\textsc{gr}}
-f{{k^\prime}_{(a}}^c\mathcal{J}^{\textsc{gr}}_{[b)c]}.
\end{gather}
\end{subequations}
The constraint algebra for this primed system consists of~\eqref{eq:Jays} combined with 
\begin{subequations}\label{eq:vacuum primed algebra}
\begin{gather}
\label{eq:vpa first}
\bigl\{\underline{\boldsymbol{\mathcal{H}}}^\prime[f],
{\underline{\boldsymbol{\mathcal{J}}}^a}_{b}[{\omega^b}_{a}]\bigr\}=0,\quad
\bigl\{\underline{\boldsymbol{\mathcal{H}}}^\prime_a[N^a],
{\underline{\boldsymbol{\mathcal{J}}}^a}_{b}[{\omega^b}_{a}]\bigr\}
=\int_\Sigma d^3x\,
N^a\Delta_{\tilde{\omega}}[\underline{\boldsymbol{\mathcal{H}}}^\prime]_a,\\
\label{eq:vpa second}
\bigl\{{\underline{\boldsymbol{\mathcal{H}}}^\prime}[f],
{\underline{\boldsymbol{\mathcal{H}}}^\prime}[g]\bigr\}
=\int_\Sigma d^3x\,(f\nabla_a[g]-g\nabla_a[f])
\bigl(\gamma^{ab}{\underline{\boldsymbol{\mathcal{H}}}^\prime}_b
-E\nabla_b[\boldsymbol{\mathcal{J}}]^{[ab]}\bigr),\\
\label{eq:vpa third}
\bigl\{{\underline{\boldsymbol{\mathcal{H}}}^\prime}[f],
{\underline{\boldsymbol{\mathcal{H}}}^\prime}_a[g^a]\bigr\}
=\int_\Sigma d^3x\,\Bigl(
\underline{f}\pounds_{\vec{g}}[{\boldsymbol{\mathcal{H}}^\prime}]
-fg^a\Delta_{\tilde{k}^\prime}
[{\underline{\boldsymbol{\mathcal{H}}}^\prime}]_a
+2g^a\nabla_c[f{k^\prime}_{ab}]
\underline{\boldsymbol{\mathcal{J}}}^{[bc]}\Bigr),\\
\label{eq:vpa fourth}
\bigl\{{\underline{\boldsymbol{\mathcal{H}}}^\prime}_a[f^a],
{\underline{\boldsymbol{\mathcal{H}}}^\prime}_b[g^b]\bigr\}
=-\int_\Sigma d^3x\, 
f^ag^b{R^c}_{dab}{\underline{\boldsymbol{\mathcal{J}}}^d}_{c},
\end{gather}
\end{subequations}
which was also derived in~\cite{Clayton:1996b}.

As we shall see, including matter presents no great difficulties.
In Section~\ref{sect:scalar} we consider a scalar field which, since it is chosen to transform trivially under frame rotations, is a brief demonstration of how matter fields fit into the generalized structure.
In Section~\ref{sect:Maxwell}, a one-form (gauge) field is introduced, which transforms  non-trivially under spatial frame rotations.
In addition, a constraint which generates $\mathrm{U}(1)$ transformations on the Cauchy surface appears, resulting in an extended diffeomorphism algebra.
Here we find a nontrivial mixing of the Maxwell canonical pair with that of the spatial vierbein---mixing that, at least superficially, resembles derivative-coupling.
We will see that this mixing is merely a manifestation of the gauge choice and is not ``true'' derivative-coupling since the metric compatibility conditions are not altered.
Finally, in Section~\ref{sect:Dirac}, the spatial frame is constrained to be orthogonal and a Dirac field is introduced.

In all three cases the matter action is added to that of vacuum GR: $S^{\textsc{gr}}\rightarrow S^{\textsc{gr}}+S^{\textsc{m}}$, and since both sets of vacuum GR constraints $(\mathcal{H}^{\textsc{gr}},\mathcal{H}^{\textsc{gr}}_a)$ and $(\mathcal{H}^{\prime\textsc{gr}},\mathcal{H}^{\prime\textsc{gr}}_a)$ are equivalent to $(-2{G^\perp}_\perp, -2{G^\perp}_a)$ ($G_{AB}$ is the Einstein tensor), we shall see that the additively combined vacuum GR and matter constraints are equivalent to adding $({T^\perp}_\perp,{T^\perp}_a)$ to these.
The matter stress-energy tensor is conveniently derived from the matter action $S^{\textsc{m}}$ via
\begin{equation}\label{eq:SE tensor}
T_{AB}=\frac{2}{E}\Biggl(
\frac{\delta S^{\textsc{m}}}{\delta\mathbf{g}^{AB}}
-\frac{1}{2}\mathrm{g}_{AB}\mathrm{g}^{CD}
\frac{\delta S^{\textsc{m}}}{\delta\mathbf{g}^{CD}}
\Biggr)
=\frac{1}{\sqrt{-\mathrm{g}}}\Biggl(
\mathrm{g}_{AC}{E^\mu}_B\frac{\delta S^{\textsc{m}}}
{\delta{\underline{E}^\mu}_C}
-\mathrm{g}_{AB}{E^\mu}_C
\frac{\delta S^{\textsc{m}}}
{\delta{\underline{E}^\mu}_C}
\Biggr);
\end{equation}
the equivalence of these two forms is a basic consistency requirement for the generalized treatment considered herein.
By an argument of Floreanini and Percacci~\cite{Floreanini+Percacci:1990} we know that once the matter action has been properly written in a general linear frame, these forms will be consistent.
Note however that in~\eqref{eq:SE tensor} we are treating the metric and vierbein degrees of freedom in the matter action \textit{independently}, which will \textit{not} work with the E-D system where only the second form is applicable.
We will show (briefly) that the matter field equations and the stress-energy tensor determined from both forms of~\eqref{eq:SE tensor} are properly generated via a variational principle, and in the canonical picture that the action of the constraints on phase space properly generates spatial diffeomorphisms once matter has been incorporated into the canonical formalism, and~\eqref{eq:vacuum unprimed} and~\eqref{eq:vacuum primed} are properly extended to reproduce~\eqref{eq:unprimed EE} and~\eqref{eq:primed EE} respectively with the appropriate stress-energy tensor.

One of the unfortunate aspects and at the same time strengths of diffeomorphism invariant theories is the great arbitrariness in parameterizing configuration and momentum space fields; any ``all-encompassing'' formalism would have to include canonical transformations that relate different parameterizations.
The content of the `geometric' derivations of the constraint algebras~\cite{Teitelboim:1973,Hojman+Kuchar+Teitelboim:1976,Clayton:1996b} is that the algebra is fixed by the reduction of the spacetime equations to the evolution of spatial quantities; that is, the algebra is a reflection of the spacetime diffeomorphism invariance.
The details of the algebra will change depending on the parameterization of phase space and the choice of constraints, however are always of a fixed form once these choices are made.
Thus the point of this work is \textit{not} to advocate any particular parameterization, merely to make it clear that there is actually a wider range of possibilities than heretofore considered.
Within this wider class one finds that the Einstein-Dirac system may be treated in a particular limit, and not as an awkward construction that seemingly must be invented solely to deal with Fermion fields.
Thus we find that the Einstein-Dirac (E-D) system is operationally no different than the Einstein--Klein Gordon (E-KG) or Einstein-Maxwell (E-M) systems; there is no more derivative-coupling in the E-D system than in the E-M system.
Indeed the mixing of canonical variables is necessary in order to retrieve the correct surface frame transformation generators, while retaining the metric compatibility conditions at the level of the field equations.

\section{Scalar Fields}
\label{sect:scalar}

The scalar field example that we consider in this section is a trivial extension of the vacuum GR results, included for completeness and to give a straightforward example of how the extension of the results of~\cite{Clayton:1997c} outlined in Section~\ref{sect:Intro} to ``GR$+$matter'' systems proceeds.
Its simplicity is due to the fact that we have chosen canonical coordinates that do not transform under changes of spatial frame.
Note that this is \textit{not} necessary; we could just as well chosen to parameterize the scalar field action in terms of a densitized scalar field, in which case some of the structure that we will encounter in Section~\ref{sect:Maxwell} would have appeared.

The scalar field Lagrangian density with self-coupling potential $V[\phi]$ is
\begin{equation}\label{eq:Scalar Lagrangian}
\begin{split}
\underline{\boldsymbol{\mathcal{L}}}^\phi
&=\boldsymbol{E}\bigl(\tfrac{1}{2}\mathrm{g}^{AB}
\pre{4}\nabla_A[\phi]\pre{4}\nabla_B[\phi]
-V[\phi]\bigr)\\
&=\frac{\boldsymbol{E}}{2N}\bigl(\partial_t[\phi]-N^ae_a[\phi]\bigr)^2
-\frac{1}{2}\underline{N}
\boldsymbol{\gamma}^{ab}\nabla_a[\phi]\nabla_b[\phi]
-\underline{\boldsymbol{N}}V[\phi],
\end{split}
\end{equation}
from which we derive (from either form of~\eqref{eq:SE tensor})
\begin{subequations}
\begin{gather}
T^\phi_{AB}
=\pre{4}\nabla_A[\phi]\pre{4}\nabla_B[\phi]
-\frac{1}{2}\mathrm{g}_{AB}\mathrm{g}^{CD}\pre{4}\nabla_C[\phi]\pre{4}\nabla_D[\phi]
+\mathrm{g}_{AB}V[\phi],\\
\bar{T}^\phi_{AB}
=\pre{4}\nabla_A[\phi]\pre{4}\nabla_B[\phi]
-\mathrm{g}_{AB}V[\phi].
\end{gather}
\end{subequations}
Using the field equations for $\phi$
\begin{equation}\label{eq:scalar field equations}
\frac{\delta S^\phi}{\delta\phi}
=E\mathbf{g}^{AB}\pre{4}\nabla_A\bigl[\pre{4}\nabla_B[\phi]\bigr]
-\boldsymbol{E}\delta_\phi[V],
\end{equation}
it is straightforward to demonstrate that the conservation law $\pre{4}\nabla_B{[T^\phi]^B}_A=0$ is satisfied.

From~\eqref{eq:Scalar Lagrangian} we see that the phase space of the scalar field is parameterized by a conjugate pair $(\phi,\underline{\boldsymbol{P}})$ where the conjugate momentum is
\begin{equation}
\underline{\boldsymbol{P}}
:=\frac{\boldsymbol{E}}{N}\bigl(\partial_t[\phi]-N^ae_a[\phi]\bigr)
=\boldsymbol{E}e_\perp[\phi].
\end{equation}
As a result, the Poisson brackets are written as $\{F,G\}=\{F,G\}_{\textsc{gr}}+\{F,G\}_{\phi}$, where $\{.,.\}_{\textsc{gr}}$ are the GR sector Poisson brackets given in~\cite{Clayton:1997c} and
\begin{equation}
\{F,G\}_{\phi}:=\int_\Sigma d^3x\,
\Biggl(
\frac{\delta F}{\delta \phi(x)}
\frac{\delta G}{\delta \underline{\boldsymbol{P}}(x)}
-\frac{\delta G}{\delta \phi(x)}
\frac{\delta F}{\delta \underline{\boldsymbol{P}}(x)}
\Biggr),
\end{equation}
is the additional contribution from the scalar field.

Since neither the extrinsic curvature nor time derivatives of the spatial metric or frame appear in the Lagrangian density~\eqref{eq:Scalar Lagrangian}, we see that the definitions of $k_{ab}$ and $k^\prime_{ab}$ in~\eqref{eq:k defs} are undisturbed and the frame rotation generators are identical to those in vacuum GR: ${\underline{\boldsymbol{\mathcal{J}}}^a}_{b}={{\underline{\boldsymbol{\mathcal{J}}}^{\textsc{gr}}}^a}_{b}$.
Thus we find that 
\begin{subequations}
\begin{align}
\underline{\boldsymbol{\mathcal{H}}}^\phi&=
\underline{\boldsymbol{\mathcal{H}}}^{\prime\phi}=
\tfrac{1}{2}\boldsymbol{E}(P_\phi)^2
+\tfrac{1}{2}E\boldsymbol{\gamma}^{ab}
\nabla_a[\phi]\nabla_b[\phi]
+\boldsymbol{E}V[\phi]
={{\underline{\boldsymbol{T}}^\phi}^\perp}_\perp,\\
\underline{\boldsymbol{\mathcal{H}}}^\phi_a&=
\underline{\boldsymbol{\mathcal{H}}}^{\prime\phi}_a=
\underline{\boldsymbol{P}}_\phi\nabla_a[\phi]
={{\underline{\boldsymbol{T}}^\phi}^\perp}_a,
\end{align}
\end{subequations}
are added to the constraints of GR.

Hamilton's equations for the scalar field 
\begin{equation}
\{\phi,H\}=NP_\phi+N^a\nabla_a[\phi],\quad
\{\underline{\boldsymbol{P}}_\phi,H\}=
E\nabla_a\bigl[N\boldsymbol{\gamma}^{ab}\nabla_b[\phi]\bigr]
-\underline{\boldsymbol{N}}\delta_\phi[V]
+E\nabla_a[N^a\boldsymbol{P}_\phi],
\end{equation}
are equivalent to~\eqref{eq:scalar field equations}, and noting that 
$\bigl\{\boldsymbol{\gamma}^{ab},
\underline{\boldsymbol{\mathcal{H}}}^{\textsc{m}}[f]\bigr\}
=\bigl\{\boldsymbol{\gamma}^{ab},
\underline{\boldsymbol{\mathcal{H}}}^{\textsc{m}}_a[f^a]\bigr\}=0$ and 
$\bigl\{{\underline{E}^i}_a,
\underline{\boldsymbol{\mathcal{H}}}^{\textsc{m}}[f]\bigr\}
=\bigl\{{\underline{E}^i}_a,
\underline{\boldsymbol{\mathcal{H}}}^{\textsc{m}}_a[f^a]\bigr\}=0$, we see that neither of~\eqref{eq:unp first} nor~\eqref{eq:pr first} are altered.
Using (with similar definitions with primes for Poisson brackets of the primed constraints) 
\begin{subequations}\label{eq:definitions of things}
\begin{align}
\bigl\{{\boldsymbol{p}^a}_i,
\underline{\boldsymbol{\mathcal{H}}}^{\phi}[f]\bigr\}
&={\boldsymbol{Y}^a}_b[f]{E^b}_i-\tfrac{1}{2}\boldsymbol{Y}[f]{E^a}_i,&
\bigl\{\underline{\pi}_{ab},
\underline{\boldsymbol{\mathcal{H}}}^{\phi}[f]\bigr\}
&=\underline{X}_{ab}[f]-\gamma_{ab}\underline{X}[f],\\
\bigl\{{\boldsymbol{p}^a}_i,
\underline{\boldsymbol{\mathcal{H}}}^{\phi}_b[f^b]\bigr\}
&={\boldsymbol{W}^a}_b[\vec{f}]{E^b}_i
-\tfrac{1}{2}\boldsymbol{W}[\vec{f}]{E^a}_i,&
\bigl\{\underline{\pi}_{ab},
\underline{\boldsymbol{\mathcal{H}}}^{\phi}_c[f^c]\bigr\}
&=\underline{Z}_{ab}[\vec{f}]-\gamma_{ab}\underline{Z}[\vec{f}],
\end{align}
\end{subequations}
where
\begin{subequations}
\begin{align}
{W^a}_b[\vec{f}]&={{W^\prime}^a}_b[\vec{f}]
=-f^a\mathcal{H}^\phi_b,\quad
Z_{ab}[\vec{f}]={Z^\prime}_{ab}[\vec{f}]=0,\\
X_{ab}[f]&=X^\prime_{ab}[f]
=-\tfrac{1}{2}fT^\phi_{ab}
=-\tfrac{1}{2}f(\bar{T}_{ab}-\gamma_{ab}\bar{T})
-\tfrac{1}{2}f\gamma_{ab}\mathcal{H}^\phi,\\
Y_{ab}[f]&={Y^\prime}_{ab}[f]
=-fT^\phi_{ab}
=-f(\bar{T}_{ab}-\gamma_{ab}\bar{T})
-f\gamma_{ab}\mathcal{H}^\phi,
\end{align}
\end{subequations}
it is straightforward to show that the evolution of the extrinsic curvature is correctly generated.
That is, we find that $\mathcal{H}^{\textsc{gr}}\rightarrow\mathcal{H}^{\textsc{gr}}+\mathcal{H}^{\phi}$ and $\tfrac{1}{2}f\bar{T}^\phi_{ab}$ is appended to the right hand sides of~\eqref{eq:unp second},~\eqref{eq:unp third},~\eqref{eq:pr second}, and~\eqref{eq:pr third} in accordance with~\eqref{eq:unprimed EE} and~\eqref{eq:primed EE}.
The unprimed and primed constraint algebras are identical to~($44$) and~($49$) respectively of~\cite{Clayton:1997c} with $\mathcal{H}=\mathcal{H}^{\textsc{gr}}+\mathcal{H}^\phi$ and $\mathcal{H}_a=\mathcal{H}^{\textsc{gr}}_a+\mathcal{H}^\phi_a$; or, equivalently,~\eqref{eq:EM-unprimed} and~\eqref{eq:EM-primed} combined with $\mathcal{H}^{\textsc{a}}_{\mathrm{U}(1)}=0$.

\section{Maxwell Fields}
\label{sect:Maxwell}

The Einstein-Maxwell system is more involved.
As the vector potential is a one-form field, it is associated to $GL\mathbf{M}$ through a (co-)vector representation of $\mathrm{GL}(4,\mathbb{R})$, and therefore results in a nontrivial contribution to ${\mathcal{J}^a}_b$.
This means that the primed and unprimed cases may no longer be considered together; the unprimed system is considered in Section~\ref{sect:unprimed} and the primed in Section~\ref{sect:primed}.
We shall see that although computationally involved, the results are a fairly straightforward extension of those of vacuum GR.

The components of the field strength tensor $F:=\mathrm{d}A=\tfrac{1}{2}F_{AB}\theta^A\wedge\theta^B$ are derived from the one-form $A=A_A\theta^A$ by
\begin{equation}\label{eq:field strength tensor}
F_{AB}=\pre{4}\nabla_A[A]_B-\pre{4}\nabla_B[A]_A
=e_A[A_B]-e_B[A_A]-{C_{AB}}^CA_C, 
\end{equation}
and the standard Lagrangian density is
\begin{equation}\label{eq:Maxwell Lagrangian}
\underline{\boldsymbol{\mathcal{L}}}^{\textsc{a}}
=-\tfrac{1}{4}\boldsymbol{E}\mathrm{g}^{AC}\mathrm{g}^{BD}F_{AB}F_{CD}
=\underline{\boldsymbol{N}}\bigl(
\tfrac{1}{2}\gamma^{ab}F_{\perp a}F_{\perp b}
-\tfrac{1}{4}F_{ab}F^{ab}\bigr),
\end{equation}
where $F_{ab}=\nabla_aA_b-\nabla_bA_a$ and
\begin{equation}\label{eq:reduce defn}
F_{\perp a}
=(\partial_t[A_a]-A_b{E^b}_i\partial_t[{E^i}_a]-\pounds_{\vec{N}}[A]_a)/N
-\nabla_a[A_\perp]-A_\perp\nabla_a[\ln(N)].
\end{equation}
The Maxwell field equations derived from~\eqref{eq:Maxwell Lagrangian} are
\begin{equation}\label{eq:Max eqns}
\frac{\delta S^{\textsc{m}}}{\delta A_A}=\boldsymbol{E}\pre{4}\nabla_B[F]^{BA}=0.
\end{equation}
Making use of the variation of the field strength tensor with respect to the spatial frame:
\begin{equation}\label{eq:delta E FST}
\delta_E F_{AB}=2\pre{4}\nabla_C[A]_{[B}{E^C}_\mu\delta{E^\mu}_{A]}
+2A_C\pre{4}\nabla_{[B}[{E^C}_\mu\delta{E^\mu}_{A]}],
\end{equation}
we find that the stress-energy tensor as determined by either form of~\eqref{eq:SE tensor} is
\begin{equation}
T^{\textsc{a}}_{AB}
=\bar{T}^{\textsc{a}}_{AB}
=-\mathrm{g}^{CD}F_{AC}F_{BD}
+\tfrac{1}{4}\mathrm{g}_{AB}F^{BC}F_{BC}.
\end{equation}
From the field equations~\eqref{eq:Max eqns} and using the Bianchi identity $\pre{4}\nabla_A[F]_{BC}+\pre{4}\nabla_C[F]_{AB}+\pre{4}\nabla_B[F]_{CA}=0$, we find that the conservation laws $\pre{4}\nabla_B{[T^{\textsc{a}}]^B}_{A}=0$ are satisfied.
Note that despite the presence of the structure constants ${C_{AB}}^C$ in the field strength tensor~\eqref{eq:field strength tensor}, the variation of the action~\eqref{eq:Maxwell Lagrangian} using~\eqref{eq:delta E FST} does \textit{not} lead to any terms containing the second-derivative of the frame.

We see from~\eqref{eq:Maxwell Lagrangian} and~\eqref{eq:reduce defn} that the momenta conjugate to $A_a$ are determined as
\begin{equation}
\frac{\delta L^{\textsc{a}}}{\delta \partial_t[A_a]}
=\underline{\boldsymbol{\gamma}}^{ab}F_{\perp b}
=:\underline{\boldsymbol{P}}^a.
\end{equation}
Therefore the phase space of the Maxwell field is parameterized by the conjugate pair $(A_a,\underline{\boldsymbol{P}})$, and the Poisson brackets are extended by (summation over $a$ is assumed)
\begin{equation}
\{F,G\}_{\textsc{a}}:=\int_\Sigma d^3x\,
\Biggl(
\frac{\delta F}{\delta A_a(x)}
\frac{\delta G}{\delta \underline{\boldsymbol{P}}^a(x)}
-\frac{\delta G}{\delta A_a(x)}
\frac{\delta F}{\delta \underline{\boldsymbol{P}}^a(x)}
\Biggr).
\end{equation}
However we also find that the definition of the momenta conjugate to ${\underline{E}^i}_a$ as given in equation~$(30)$ of~\cite{Clayton:1997c} is extended to
\begin{equation}
{\boldsymbol{p}^a}_{i}:=-2{\boldsymbol{K}^a}_{b}{E^b}_i
-\boldsymbol{P}^aA_b{E^b}_i
+\tfrac{1}{2}A^b\boldsymbol{P}_b{E^a}_i,
\end{equation}
which is the ``geometrodynamical momentum'' conjugate to the frame, the first term of which is the ``gravitational momentum''~\cite{Kuchar:1977}.
Thus we find that this definition mixes the gravitational and non-gravitational fields, and DeWitt's super-metric is no longer block-diagonal.
By defining
\begin{equation}
\kappa_{ab}:=-\tfrac{1}{2}(P_{(a}A_{b)}-\tfrac{1}{2}\gamma_{ab}P^cA_c),
\end{equation}
we now have (\textit{c.f.} the discussion following~($33$) of~\cite{Clayton:1997c})
\begin{equation}
{K}_{ab}\approx
\bar{a}k_{ab}
+(1-\bar{a})(k^\prime_{ab}+\kappa_{ab}),
\end{equation}
where the unprimed case corresponds to $\bar{a}=1$ and the primed case to $\bar{a}=0$.
In addition, the generators of frame transformations become
\begin{equation}
{\underline{\boldsymbol{\mathcal{J}}}^a}_{b}=
2\boldsymbol{\gamma}^{ac}\underline{\pi}_{bc}
-\underline{\boldsymbol{\pi}}\delta^a_b
-{\boldsymbol{p}^a}_i{\underline{E}^i}_b
+\underline{\boldsymbol{p}}\delta^a_b
-\underline{\boldsymbol{P}}^aA_b,
\end{equation}
which correctly generate frame transformations on the vector field sector
\begin{equation}
\bigl\{A_a,{\underline{\boldsymbol{\mathcal{J}}}^b}_{c}[{\omega^c}_{b}]\bigr\}
=\Delta_{\tilde{\omega}}[A]_a,\quad
\bigl\{\underline{\boldsymbol{P}}^a,
{\underline{\boldsymbol{\mathcal{J}}}^b}_{c}[{\omega^c}_{b}]\bigr\}
=\Delta_{\tilde{\omega}}[\underline{\boldsymbol{P}}]^a,
\end{equation}
and satisfy~\eqref{eq:Jays}.

The Lagrange multiplier $A_\perp$ enforces the $\mathrm{U}(1)$ constraint:
\begin{equation}\label{eq:U1 constraint}
\underline{\boldsymbol{\mathcal{H}}}^{\textsc{a}}_{\mathrm{U}(1)}
=-\boldsymbol{E}\nabla_a[P]^a,
\end{equation}
and we can immediately compute
\begin{equation}\label{eq:YOUJAY}
\bigl\{\underline{\boldsymbol{\mathcal{H}}}^{\textsc{a}}_{\mathrm{U}(1)}
[\alpha],
\underline{\boldsymbol{\mathcal{H}}}^{\textsc{a}}_{\mathrm{U}(1)}
[\beta]\bigr\}
=0,\quad
\bigl\{\underline{\boldsymbol{\mathcal{H}}}^{\textsc{a}}_{\mathrm{U}(1)}
[\alpha],
{\underline{\boldsymbol{\mathcal{J}}}^a}_{b}[{\omega^b}_{a}]
\bigr\}
=0,
\end{equation}
and the non-vanishing brackets
\begin{subequations}\label{eq:U1 brackets}
\begin{gather}
\bigl\{A_a,\underline{\boldsymbol{\mathcal{H}}}^{\textsc{a}}_{\mathrm{U}(1)}
[\alpha]\bigr\}
=\nabla_a[\alpha],\quad
\bigl\{\underline{\boldsymbol{P}}^a,
\underline{\boldsymbol{\mathcal{H}}}^{\textsc{a}}_{\mathrm{U}(1)}
[\alpha]\bigr\}=0,\\
\bigl\{{\boldsymbol{p}^a}_i,
\underline{\boldsymbol{\mathcal{H}}}^{\textsc{a}}_{\mathrm{U}(1)}
[\alpha]\bigr\}
=-(\boldsymbol{P}^a{E^b}_i-\tfrac{1}{2}\boldsymbol{P}^b{E^a}_i)\nabla_b[\alpha].
\end{gather}
\end{subequations}
This additional constraint is appended to the standard forms of the Hamiltonian via an additional atlas field $\alpha:=NA_\perp$ as (and similarly for the primed system)
\begin{align}\label{eq:STFs}
H^{\textsc{a}}&=\int_\Sigma d^3x\,\bigl(
N\underline{\boldsymbol{\mathcal{H}}}
+N^a\underline{\boldsymbol{\mathcal{H}}}_a
+{N^a}_b{\underline{\boldsymbol{\mathcal{J}}}^b}_{a}
+\alpha\underline{\boldsymbol{\mathcal{H}}}^{\textsc{a}}_{\mathrm{U}(1)}
\bigr).
\end{align}
To determine the form of the other constraints appearing in~\eqref{eq:STFs} we follow the development in~\cite{Clayton:1997c}: choosing $K_{ab}=k_{ab}$ ($\bar{a}=1$) leads to the unprimed system that closely resembles the coordinate frame approach (Section~\ref{sect:unprimed}), whereas choosing $K_{ab}=k^\prime_{ab}+\kappa_{ab}$ ($\bar{a}=0$) leads to the primed case (Section~\ref{sect:primed}).

\subsection{The Unprimed System}
\label{sect:unprimed}

Since for the unprimed system we have $K_{ab}=k_{ab}$ the results are straightforward to derive, merely pulling off the Hamiltonian and momentum constraints from the Maxwell Hamiltonian, resulting in the following additions to the unprimed constraints
\begin{subequations}\label{eq:Max unprimed constraints}
\begin{align}
\underline{\boldsymbol{\mathcal{H}}}^{\textsc{a}}:&=
\tfrac{1}{2}\boldsymbol{E} P^a P_a
+\tfrac{1}{4}\boldsymbol{E}F^{ab}F_{ab}
={{\underline{\boldsymbol{T}}^{\textsc{a}}}^\perp}_\perp,\\
\underline{\boldsymbol{\mathcal{H}}}^{\textsc{a}}_a:
&=\underline{\boldsymbol{P}}^bF_{ab}
-\underline{\boldsymbol{A}}_a\nabla_b[P]^b
={{\underline{\boldsymbol{T}}^{\textsc{a}}}^\perp}_a
+A_a\underline{\boldsymbol{\mathcal{H}}}^{\textsc{a}}_{\mathrm{U}(1)}.
\end{align}
\end{subequations}
From these it is straightforward to determine
\begin{subequations}\label{eq:unprimed Max}
\begin{align}
\bigl\{A_a,\underline{\boldsymbol{\mathcal{H}}}^{\textsc{a}}[f]\bigr\}
&=fP_a,&
\bigl\{\underline{\boldsymbol{P}}^a,
\underline{\boldsymbol{\mathcal{H}}}^{\textsc{a}}[f]\bigr\}
&=-\boldsymbol{E}\nabla_b[fF^{ab}],\\
\bigl\{A_a,\underline{\boldsymbol{\mathcal{H}}}_b[f^b]\bigr\}
&=\pounds_{\vec{f}}[A]_a,&
\bigl\{\underline{\boldsymbol{P}}^a,
\underline{\boldsymbol{\mathcal{H}}}_b[f^b]\bigr\}
&=E\pounds_{\vec{f}}[\boldsymbol{P}]^a,
\end{align}
\end{subequations}
and using~\eqref{eq:U1 brackets}, Hamilton's equations for the Maxwell field are found to be
\begin{subequations}\label{eq:unprimed Ham Max}
\begin{align}
\bigl\{A_a,H\bigr\}&=N P_a+\pounds_{\vec{N}}[A]_a
+\Delta_{\tilde{N}}[A]_a+\nabla_a[\alpha],\\
\bigl\{\underline{\boldsymbol{P}}^a,H\bigr\}
&=-\boldsymbol{E}\nabla_b[NF^{ab}]
+E\pounds_{\vec{N}}[\boldsymbol{P}]^a
+\Delta_{\tilde{N}}[\underline{\boldsymbol{P}}]^a,
\end{align}
\end{subequations}
which, when combined combined with the $\mathrm{U}(1)$ constraint~\eqref{eq:U1 constraint}, are equivalent to~\eqref{eq:Max eqns}.

Using the definitions as given in~\eqref{eq:definitions of things}, we find
\begin{subequations}
\begin{align}
{\boldsymbol{W}^a}_{b}[\vec{f}]
&=-\pounds_{\vec{f}}[A_b\boldsymbol{P}^a]
-f^a\underline{\boldsymbol{\mathcal{H}}}^{\textsc{a}}_b,&
Z_{ab}&=0,\\
{Y^a}_b[f]&=
-f\bar{T}^{\textsc{a}}_{ab}
-fP_aP_b
-A_b\nabla_c[fF^{ca}],&
X_{ab}[f]&
=-\tfrac{1}{2}f\bar{T}^{\textsc{a}}_{ab}.
\end{align}
\end{subequations}
Noting that $\bar{T}:=\gamma^{ab}\bar{T}_{ab}=\mathcal{H}^{\textsc{a}}$, it is straightforward to show that $\{k_{ab},\underline{\boldsymbol{\mathcal{H}}}\}=\tfrac{1}{2}f\bar{T}_{ab}-\tfrac{1}{4}\gamma_{ab}\mathcal{H}^{\textsc{a}}$, and therefore the correct extension of~\eqref{eq:unp second} is generated.
In order to check the extension of~\eqref{eq:unp third},  note that we now must compute $\{k^\prime_{ab}+\kappa_{ab},\underline{\boldsymbol{\mathcal{H}}}[f]\}$ which correctly results in the additional contributions $\tfrac{1}{2}f\bar{T}_{ab} -\tfrac{1}{4}f\gamma_{ab}\mathcal{H}^{\textsc{a}} -fk \kappa_{ab} +2f{k_{(a}}^c \kappa_{cb)} +f{k_{(a}}^c\mathcal{J}^{\textsc{a}}_{[cb)]}$. 

It is now possible to compute the algebra of the unprimed constraints, finding~\eqref{eq:Jays},~\eqref{eq:YOUJAY},~\eqref{eq:vua first},~\eqref{eq:vua second} and 
\begin{subequations}\label{eq:EM-unprimed}
\begin{gather}
\bigl\{\underline{\boldsymbol{\mathcal{H}}}^{\textsc{a}}_{\mathrm{U}(1)}
[\alpha],
\underline{\boldsymbol{\mathcal{H}}}[f]
\bigr\}=0,\quad
\bigl\{\underline{\boldsymbol{\mathcal{H}}}^{\textsc{a}}_{\mathrm{U}(1)}
[\alpha],
\underline{\boldsymbol{\mathcal{H}}}_a[f^a]
\bigr\}
=-\int_\Sigma d^3x\,\underline{\alpha}\pounds_{\vec{f}}
[\boldsymbol{\mathcal{H}}^{\textsc{a}}_{\mathrm{U}(1)}],\\
\bigl\{\underline{\boldsymbol{\mathcal{H}}}[f],
\underline{\boldsymbol{\mathcal{H}}}[g]\bigr\}
=(f\nabla_a[g]-g\nabla_a[f])\gamma^{ab}(
\underline{\boldsymbol{\mathcal{H}}}_b
-A_b\underline{\boldsymbol{\mathcal{H}}}^{\textsc{a}}_{\mathrm{U}(1)}).
\end{gather}
\end{subequations}
Note that the presence of the $\mathrm{U}(1)$ constraint is a new feature of the algebra not anticipated by the analysis of~\cite{Clayton:1996b}.

\subsection{The Primed System}
\label{sect:primed}

The primed constraints are
\begin{subequations}
\begin{align}
{\underline{\boldsymbol{\mathcal{H}}}^\prime}^{\textsc{a}}
&=\tfrac{1}{2}\boldsymbol{E}P^aP_a
+\tfrac{1}{4}\boldsymbol{E}F^{ab}F_{ab}
+({\kappa^a}_b{\kappa^b}_a-\kappa^2)
+2({\kappa^a}_b{{k^\prime}^b}_a-k^\prime\kappa),\\
{\underline{\boldsymbol{\mathcal{H}}}^\prime}^{\textsc{a}}_a
&=\underline{\boldsymbol{P}}^b\nabla_a[A]_b,
\end{align}
\end{subequations}
where note that they are related to~\eqref{eq:Max unprimed constraints} via combinations of the frame rotation generators as given in ($45$) of~\cite{Clayton:1997c}, and therefore the combined constraints are equivalent to the Einstein equations ${G^\perp}_A=\tfrac{1}{2}{T^\perp}_A$.
Their action on the Maxwell phase space is found to be
\begin{subequations}\label{eq:primed Max}
\begin{align}
\bigl\{A_a,
{\underline{\boldsymbol{\mathcal{H}}}^\prime}^{\textsc{a}}[f]\bigr\}
&=fP_a+f\Delta_{\tilde{k}^\prime+\tilde{\kappa}}[A]_a,&
\bigl\{\underline{\boldsymbol{P}}^a,
{\underline{\boldsymbol{\mathcal{H}}}^\prime}^{\textsc{a}}[f]\bigr\}
&=-\boldsymbol{E}\nabla_b[fF^{ab}]
+f\Delta_{\tilde{k}^\prime+\tilde{\kappa}}[\underline{\boldsymbol{P}}]^a,\\
\bigl\{A_a,
{\underline{\boldsymbol{\mathcal{H}}}^\prime}^{\textsc{a}}_a[f^a]\bigr\}
&=f^b\nabla_b[A]_a=\pounds^\prime_{\vec{f}}[A]_a,&
\bigl\{\underline{\boldsymbol{P}}^a,
{\underline{\boldsymbol{\mathcal{H}}}^\prime}^{\textsc{a}}_a[f^a]\bigr\}
&=\boldsymbol{E}\nabla_b[f^bP^a]
=\pounds^\prime_{\vec{f}}[\underline{\boldsymbol{P}}]^a.
\end{align}
\end{subequations}
Note that the actions~\eqref{eq:unprimed Max} and~\eqref{eq:primed Max} are those of $\mathrm{d}_\perp$ and $\pounds$ and $\mathrm{d}^\prime_\perp$ and $\pounds^\prime$ respectively.
Hamilton's equations for the Maxwell sector are straightforward to find, and, as in the case of vacuum GR (see~($52$) of~\cite{Clayton:1997c}), are weakly equivalent to~\eqref{eq:unprimed Ham Max} when one makes the replacements $N=N^\prime$, $N_a=N^\prime_a$, ${N^a}_b={{N^\prime}^a}_b+\nabla_b[N^\prime]^a+N^\prime{k^a}_b$.

Because we find $\bigl\{{\underline{E}^i}_a,
{\underline{\boldsymbol{\mathcal{H}}}^\prime}^{\textsc{a}}[f]\bigr\}
=\Delta_{\tilde{\kappa}}[{\underline{E}^i}_a]$, 
the action of the Hamiltonian constraint on the frame becomes
\begin{equation}
\bigl\{{\underline{E}^i}_a,
{\underline{\boldsymbol{\mathcal{H}}}^\prime}[f]\bigr\}
=f\Delta_{\tilde{k}^\prime+\tilde{\kappa}}[{\underline{E}^i}_a],
\end{equation}
properly reproducing the evolution of the vierbein in accordance with~\eqref{eq:prime evol comp}.
Again making use of the definitions~\eqref{eq:definitions of things} we find
\begin{subequations}
\begin{align}
W^\prime_{ab}[\vec{f}]&=
-f_a{\mathcal{H}^\prime}^{\textsc{a}}_b
+\nabla_c[f^cP_{[b}A_{a]}]
+\nabla_c[P^cA_{(b}f_{a)}]
-\nabla_c[A^cf_{(b}P_{a)}],\\
X^\prime_{ab}[f]&=
-\tfrac{1}{2}f\bar{T}^{\textsc{a}}_{ab}
+\tfrac{1}{2}f\gamma_{ab}({\kappa^a}_b{\kappa^b}_a-\kappa^2)
+f\gamma_{ab}({{k^\prime}^a}_b{\kappa^b}_a-k^\prime\kappa)\nonumber \\
&+f{\kappa_{(a}}^c\mathcal{J}^{\textsc{gr}}_{[b)c]}
+f({{k^\prime}_{(a}}^c+{\kappa_{(a}}^c)\mathcal{J}^{\textsc{a}}_{[b)c]},\\
Y^\prime_{ab}[f]&=
-f\bar{T}_{ab}
-fP_aP_b
+A_b\nabla_c[f{F_a}^c]\nonumber \\
&-2f({\kappa_a}^ck^\prime_{bc}-\kappa k^\prime_{ab})
-f\gamma_{ab}({\kappa^a}_b{\kappa^b}_a-\kappa^2)
+f({\kappa_a}^c-\delta^c_a\kappa)\mathcal{J}^{\textsc{gr}}_{[bc]},\\
Z^\prime_{ab}[\vec{f}]&=
\tfrac{1}{2}\bigl(
\nabla_c[P^cf_{(a}A_{b)}]
+\nabla_c[f^cP_{(a}A_{b)}]
-\nabla_c[A^cf_{(a}P_{b)}]
\bigr),
\end{align}
\end{subequations}
and noting that ${\kappa_{(a}}^c\mathcal{J}^{\textsc{a}}_{[cb)]}
=\tfrac{1}{8}P^cP_c A_a A_b
-\tfrac{1}{8}A^cA_c P_a P_b$, it is straightforward to show that the appropriate additions to~\eqref{eq:pr second} and~\eqref{eq:pr third} are generated from these.

The primed constraint algebra consists of~\eqref{eq:Jays},~\eqref{eq:YOUJAY},\eqref{eq:vpa first},~\eqref{eq:vpa fourth} and
\begin{subequations}\label{eq:EM-primed}
\begin{gather}
\bigl\{\underline{\boldsymbol{\mathcal{H}}}^{\textsc{a}}_{\mathrm{U}(1)}
[\alpha],
\underline{\boldsymbol{\mathcal{H}}}^\prime[f]
\bigr\}=0,\quad
\bigl\{\underline{\boldsymbol{\mathcal{H}}}^{\textsc{a}}_{\mathrm{U}(1)}
[\alpha],
\underline{\boldsymbol{\mathcal{H}}}^\prime_a[f^a]
\bigr\}
=-\int_\Sigma d^3x\,\alpha\pounds^\prime_{\vec{f}}
\bigl[\underline{\boldsymbol{\mathcal{H}}}^{\textsc{a}}_{\mathrm{U}(1)}\bigr],\\
\bigl\{{\underline{\boldsymbol{\mathcal{H}}}^\prime}[f],
{\underline{\boldsymbol{\mathcal{H}}}^\prime}[g]\bigr\}
=\int_\Sigma d^3x\,(f\nabla_a[g]-g\nabla_a[f])
\bigl(\gamma^{ab}{\underline{\boldsymbol{\mathcal{H}}}^\prime}_b
-E\nabla_b[\boldsymbol{\mathcal{J}}]^{[ab]}
-A^a\underline{\boldsymbol{\mathcal{H}}}^{\textsc{a}}_{\mathrm{U}(1)}\bigr),\\
\bigl\{{\underline{\boldsymbol{\mathcal{H}}}^\prime}[f],
{\underline{\boldsymbol{\mathcal{H}}}^\prime}_a[g^a]\bigr\}
=\int_\Sigma d^3x\,\Bigl(
\underline{f}\pounds_{\vec{g}}[{\boldsymbol{\mathcal{H}}^\prime}]
-fg^a\Delta_{\tilde{k}^\prime+\tilde{\kappa}}
[{\underline{\boldsymbol{\mathcal{H}}}^\prime}]_a
+2g^a\nabla_c[f({k^\prime}_{ab}+\kappa_{ab})]
\underline{\boldsymbol{\mathcal{J}}}^{[bc]}\Bigr).
\end{gather}
\end{subequations}

\section{Dirac Spinors}
\label{sect:Dirac}

In order to consider the introduction of a self-gravitating Dirac spinor, we need to do some preparatory work.
The conditions for the existence of a spin structure on a manifold are known~\cite{Geroch:1968,Clarke:1971}, however since we are primarily interested in the initial-value formalism, the result that any globally hyperbolic spacetime or spacetime that admits a Cauchy surface admits a spin structure~\cite{Geroch:1970} is the most relevant.
The general theory of spinors in curved spacetime (see~\cite{Wald:1984}) requires the introduction of an $\mathrm{SL}(2,\mathbb{C})$ (the universal covering group of the Lorentz group) principle bundle, and a Dirac spinor is associated to it through the product of a vector and conjugate vector representations of $\mathrm{SL}(2,\mathbb{C})$.
Generally one introduces spin frames, and we would then need to develop formalism to deal with the initial-value problem~\cite{Sen:1982}.

Here we will pursue a more straightforward, operationally-oriented path.
We note that a Dirac spinor is associated directly to the Lorentz group through a spinor representation~\cite{Nakahara:1990}, and so we will reduce the general linear frame bundle $GL\mathbf{M}$ to the Lorentz frame bundle $L\mathbf{M}$, with structure group $\mathrm{O}(1,3)$.
(In fact, for the initial value problem we will be considering time and space oriented frames and so have in effect reduced this further to $L^+_{\uparrow}\mathbf{M}$ with structure group the proper Lorentz group $\mathcal{L}^+_{\uparrow}(1,3)=\{\Lambda\in\mathrm{O}(1,3)|\det(\Lambda)=+1,{\Lambda^0}_{0}>0\}$.)
Elements of $\mathrm{GL}(4,\mathbb{R})$ are specialized to $\mathrm{O}(1,3)$ as $M^A_B\rightarrow{\Lambda^A}_B$ and $\lvert M^{-1}\rvert^A_B\rightarrow {\Lambda_B}^A =\eta^{AC}\eta_{BD}{\Lambda^D}_C$, where ${\Lambda^A}_C{\Lambda_B}^C=\delta^A_B={\Lambda^C}_B{\Lambda_C}^A$.

We assume that given a Lorentz transformation $\Lambda$, the spinor $\psi$ transforms as
\begin{equation}
\psi\rightarrow S(\Lambda)\psi,
\end{equation}
the gamma matrices that satisfy the usual Dirac relations (we make use of the definitions and identities in Appendix~2A of~\cite{Thaller:1992})
\begin{equation}
\{\gamma^A,\gamma^B\}=2\eta^{AB},
\end{equation}
and satisfy ${\Lambda^A}_B\gamma^B=S^{-1}(\Lambda)\gamma^AS(\Lambda)$.
Writing an infinitesimal Lorentz transformation as ${\Lambda^A}_B\approx \delta^A_B+{\omega^A}_B$, ${\Lambda_B}^A\approx \delta^A_B-{\omega^A}_B$, and ${\omega^A}_B$ are the antisymmetric generators of $\mathfrak{so}(1,3)$, for the spinor representation we write $S(\Lambda)\approx 1-\tfrac{1}{4}i\hat{\omega}$ where $\hat{\omega}:={\omega^A}_B{\sigma^B}_A$ and $\sigma^{AB}:=i\tfrac{1}{2}[\gamma^A,\gamma^B]$ satisfy the commutators
\begin{equation}\label{eq:sigma commutator}
[\sigma^{AB},\sigma^{CD}]=-2i\bigl(
\sigma^{AC}\eta^{BD}
-\sigma^{AD}\eta^{BC}
+\sigma^{BD}\eta^{AC}
-\sigma^{BC}\eta^{AD}\bigr).
\end{equation}
We therefore define the action of the generator of frame rotations $\Delta$ on the spinor as
\begin{equation}\label{eq:Lorentz generators}
\Delta_{\tilde{\omega}}[\psi]
=-\tfrac{1}{4}i\hat{\omega}\psi,
\end{equation}
and using~\eqref{eq:sigma commutator} it is straightforward to show that these generators satisfy the algebra
\begin{equation}
[\Delta_{\tilde{\omega}_1},\Delta_{\tilde{\omega}_2}]\psi
=-\Delta_{\tilde{[\omega_1,\omega_2]}}[\psi].
\end{equation}
The covariant derivative operates on these spinors as
\begin{equation}
\pre{4}\nabla_A[\psi]
=e_A[\psi]+\tfrac{1}{4}i\Gamma^B_{AC}{\sigma^C}_B\psi
=e_A[\psi]+\tfrac{1}{4}i\pre{4}\tilde{\Gamma}_A\psi,
\end{equation}
and we note that $[\pre{4}\nabla_{A},\gamma^B]=0$.

The curved spacetime Dirac action is (see Section~$7.10.2$ of~\cite{Nakahara:1990})
\begin{equation}\label{eq:Dirac action}
S^{\textsc{d}}=\int d^4x\,\det({E^A}_\mu)\,\bar{\psi}
\bigl(i\tfrac{1}{2}\gamma^A\pre{4}\overleftrightarrow{\nabla}_{A}-m\bigr)\psi,
\end{equation}
where we employ the standard notation $\bar{\psi}\pre{4}\overleftrightarrow{\nabla}_A\psi:=\bar{\psi}\pre{4}\nabla_A[\psi]-\pre{4}\nabla_A[\bar{\psi}]\psi$.
Variation of this with respect to $\bar{\psi}$ and $\psi$ leads to the Dirac equation
\begin{equation}\label{eq:Dirac equation}
i\gamma^A\pre{4}\nabla_{A}[\psi]-m\psi=0,
\end{equation}
and its adjoint respectively.
As a consequence, the vector $J^A:=\bar{\psi}\gamma^A\psi$ constructed from a solution to the Dirac equation is covariantly conserved: $\pre{4}\nabla_A[J]^A=0$, which in the surface-adapted frame becomes $\partial_t[E\psi^\dagger\psi] -E\nabla_a[N^a\psi^\dagger\psi] +E\nabla_a[N\psi^\dagger\alpha^a\psi]=0$ and has the usual quantum-mechanical probabilistic interpretation.
Unlike the scalar and Maxwell field examples considered previously, the spinor is \textit{not} dimensionless after one has chosen $16\pi\mathrm{G}=\mathrm{c}=1$.
Indeed, this length scale may be interpreted as the Planck length $\textsc{l}_{\textsc{p}}$, or $\hbar$.
Here we may either consider $\psi$ to have a dimension of $\mathit{length}^{-1/2}$ or to have taken $\hbar=1$.
Note also that in this section we will introduce neither $\gamma^\mu:={E^\mu}_A\gamma^A$ nor the coordinate components of the spacetime metric.

Using the variation of the `spin-connection' with respect to the vierbein
\begin{equation}
\begin{split}
\delta\hat{\Gamma}_{A}
&=\delta{E^\mu}_A{E^D}_\mu\hat{\Gamma}_{D}\\
&+\bigl(\eta_{BD}\pre{4}\nabla_{A}[{E^D}_\mu\delta{E^\mu}_C]
-\eta_{BD}\pre{4}\nabla_{C}[{E^D}_\mu\delta{E^\mu}_A]
+\eta_{AD}\pre{4}\nabla_{B}[{E^D}_\mu\delta{E^\mu}_C]\bigr)\sigma^{CB},
\end{split}
\end{equation}
and the second form of~\eqref{eq:SE tensor}, we find the stress-energy tensor
\begin{equation}\label{eq:Dirac S-E}
T^{\textsc{d}}_{AB}=i\tfrac{1}{2}\bar{\psi}\gamma_{(A}\pre{4}\overleftrightarrow{\nabla}_{B)}\psi,\quad
T^{\textsc{d}}=m\bar{\psi}\psi.
\end{equation}
Once again we find that there are no terms containing second derivatives of either the spatial metric or frame, albeit trivially because the Dirac action~\eqref{eq:Dirac action} only contains terms that are at most linear in first-order derivatives of these fields.
The conservation laws $\pre{4}\nabla_B{[T^{\textsc{d}}]^B}_A=0$ follow from the use of~\eqref{eq:Dirac equation}, the Bianchi identities of the Riemann tensor and the wave equation satisfied by the Fermion field: $(\eta^{AB}\pre{4}\nabla_A\pre{4}\nabla_B+m^2-\tfrac{1}{4}R)\psi=0$, where we have used that $[\pre{4}\nabla_A,\pre{4}\nabla_B]\psi=\tfrac{1}{4}i{R^C}_{DAB}{\sigma^D}_C\psi=\tfrac{1}{4}i\hat{R}_{AB}\psi$ and $\hat{R}_{AB}\sigma^{AB}=-2R$.

\subsection{The Initial-Value Formalism}
\label{sect:init Dirac}

To specialize the primed system of vacuum GR to consider orthonormal frames is a simple matter of replacing $\gamma_{ab}$ by $\delta_{ab}$ in all results and ignoring anything related to the $(\boldsymbol{\gamma}^{ab},\underline{\pi}_{ab})$ part of phase space (ending up with a description of canonical tetrad GR similar to that discussed in~\cite{Banados+Contreras:1998}).
In addition, it will be useful to introduce (using the conventions of Section~1.1 of~\cite{Thaller:1992}) $\beta=\gamma^0$ and $\alpha^a=\gamma^0\gamma^a$, in terms of which we have $\sigma^{0a}=i\alpha^a$, and the spin matrices $\sigma^{ab}:=-i\tfrac{1}{2}[\alpha^a,\alpha^b]$ satisfying the reduced version of~\eqref{eq:sigma commutator}
\begin{equation}\label{small sigma commutator}
[\sigma^{ab},\sigma^{cd}]
=2i(\delta^{bc}\sigma^{da}+\delta^{bd}\sigma^{ac}
+\delta^{ca}\sigma^{bd}+\delta^{da}\sigma^{cb}).
\end{equation}
We define the reduction of the operator~\eqref{eq:Lorentz generators} that generates infinitesimal $\mathrm{O}(3)$ frame rotations by
\begin{equation}
\Delta_{\tilde{\omega}}\psi=i\tfrac{1}{4}\hat{\omega}\psi,
\quad\text{where}\quad \hat{\omega}=\omega_{ab}\sigma^{ab}, 
\end{equation}
and the surface-covariant derivative as
\begin{equation}
\nabla_{a}[\psi]
=e_a[\psi]+i\tfrac{1}{4}\hat{\Gamma}_a\psi,
\end{equation}
where $\hat{\Gamma}_a:=\delta_{bd}\Gamma^b_{ac}\sigma^{bc}$.
The commutator of these derivatives acting on a spinor results in 
$[\nabla_a,\nabla_b]\psi
=\tfrac{1}{4}i\hat{R}_{ab}\psi
=\tfrac{1}{4}iR_{cdab}\sigma^{cd}\psi$.

The decomposition of the spin connection is given by 
$i\tfrac{1}{4}\pre{4}\hat{\Gamma}_{a}
=i\tfrac{1}{4}\hat{\Gamma}_a+
\tfrac{1}{2}K_{ab}\alpha^b$, and 
$i\tfrac{1}{4}\pre{4}\hat{\Gamma}_{\perp}
=\tfrac{1}{2}a_a\alpha^a
-i\tfrac{1}{4}\hat{C}_\perp$ ($\hat{C}_\perp:=\delta_{bc}{C_{\perp a}}^b\sigma^{ac}$ and ${C_{\perp a}}^b$ is given in equation~(18b) of~\cite{Clayton:1997c}, and it's role in the surface covariant normal derivative is discussed in Section~\textrm{III}-B of the same reference), so that we have
\begin{subequations}
\begin{align}
\pre{4}\nabla_{a}[\psi]
&=\nabla_{a}[\psi]+\tfrac{1}{2}K_{ab}\alpha^b\psi,\\
\pre{4}\nabla_{\perp}[\psi]
&=\tfrac{1}{N}\partial_t[\psi]
-\tfrac{1}{N}N^a\nabla_a[\psi]
-\tfrac{i}{4N}\nabla_a[N]_b\sigma^{ab}\psi
+\tfrac{1}{2}a_a\alpha^a\psi
+\tfrac{i}{4N}E_{ai}\partial_t[{E^i}_b]\sigma^{ab}\psi.
\end{align}
\end{subequations}
Using these and introducing the ``half-densitized'' spinor fields (since in this section we have $\det(\gamma_{ab})=\det(\delta_{ab})\equiv 1$, densities of this type will not appear and there should be no confusion with boldfaced quantities that appear elsewhere)
\begin{equation}\label{eq:half psi}
\boldsymbol{\psi}:=E^{1/2}\psi,\quad
\boldsymbol{\psi}^\dagger:=E^{1/2}\psi^\dagger,
\end{equation}
from~\eqref{eq:Dirac action} it is straightforward to deduce the curved spacetime Dirac Lagrangian
\begin{multline}\label{eq:Dirac Lagrangian}
L^{\textsc{d}}
=\int_\Sigma d^3x\,\bigl(
i\tfrac{1}{2}\boldsymbol{\psi}^\dagger
\overleftrightarrow{\partial}_t\boldsymbol{\psi}
+i\tfrac{1}{2}NE\psi^\dagger\alpha^a\overleftrightarrow{\nabla}_{a}\psi
-mN\boldsymbol{\psi}^\dagger\beta\boldsymbol{\psi}\\
-i\tfrac{1}{2}EN^a\psi^\dagger\overleftrightarrow{\nabla}_a\psi
+\tfrac{1}{4}\nabla_a[N]_b\boldsymbol{\psi}^\dagger\sigma^{ab}\boldsymbol{\psi}
-\tfrac{1}{4}E_{ai}\partial_t[{E^i}_b]
\boldsymbol{\psi}^\dagger\sigma^{ab}\boldsymbol{\psi}\bigr).
\end{multline}

In order to perform the Legendre transform to obtain the E-D Hamiltonian, we need to choose an appropriate set of canonical coordinates in the Fermionic sector.
In a flat spacetime this is straightforward: one finds that (discarding a total time derivative from the action) the real and imaginary parts of $\psi$ are canonically conjugate, and one can introduce complex coordinates on phase space and use the quantum-mechanical symplectic form~\cite{Marsden+Ratiu:1994} for a four component vector.
In a curved spacetime one finds that it is the densitized imaginary part of $\psi$ that is conjugate to the real part, however introducing the half-densitized spinors~\eqref{eq:half psi} allows the flat spacetime construction to proceed in an identical manner.
Using the inner product on $\Sigma$ (written first in covariant form, and then specialized to a surface-normal frame)
\begin{equation}
(\psi,\phi)_\Sigma
=\int_\Sigma d\Sigma(x)n_A\,\bar{\psi}\gamma^A\phi
=\int_\Sigma dx\,\boldsymbol{\psi}^\dagger\boldsymbol{\phi}
\end{equation}
to define the $L^2(\Sigma)$ pairing and therefore the quantum-mechanical symplectic form, the introduction of $(\boldsymbol{\psi},\boldsymbol{\psi}^\dagger)$ as complex coordinates on the Dirac sector of phase space results in the Poisson brackets
\begin{equation}\label{eq:Dirac poisson}
\{F,G\}_{\textsc{d}}=-i\int_\Sigma dx\,\Biggl(
\frac{\delta F}{\delta\boldsymbol{\psi}(x)}
\frac{\delta G}{\delta\boldsymbol{\psi}^\dagger(x)}
-\frac{\delta G}{\delta\boldsymbol{\psi}(x)}
\frac{\delta F}{\delta\boldsymbol{\psi}^\dagger(x)}
\Biggr).
\end{equation}

On a fixed gravitational background, the result of the Legendre transformation would be to drop the first term in~\eqref{eq:Dirac Lagrangian} and $H^{\textsc{d}}$ is the negative of what remains.
For the E-D system, we find that the final term in~\eqref{eq:Dirac Lagrangian} contributes to the conjugate momentum of the vierbein, leading to
\begin{equation}
{p^a}_i:=-2{K^a}_{b}{E^b}_i
+\tfrac{1}{4}\psi^\dagger\sigma^{ab}\psi E_{bi},
\end{equation}
and once again the ``geometrodynamical momentum'' is not equal to the ``gravitational momentum''.
However unlike the Maxwell case, the definition of the extrinsic curvature on phase space is undisturbed ($K_{ab}=k^\prime_{ab}$ and~\eqref{eq:pr first} is unaltered); we find only a contribution to the frame rotation generators
\begin{equation}\label{eq:Dirac rotation generators}
\mathcal{J}_{[ab]}
:=-p_{[ai}{E^i}_{b]}
+\tfrac{1}{4}\psi^\dagger\sigma_{ab}\psi.
\end{equation}
(Note that it is possible, and indeed somewhat simpler, to represent these rotation operators using $\mathcal{S}^a=\tfrac{1}{2}\epsilon^{abc}\mathcal{J}^{[bc]}$, making use of the spin matrices $s^a:=\tfrac{1}{2}\epsilon^{abc}\sigma^{bc}=\sigma^a\mathbf{1}$ where $\sigma^a$ are the Pauli spin matrices.
We do not do so here simply because we wish to present results that are as similar as possible to those of the $\mathrm{GL}(3,\mathbb{R})$ case.)
In addition to generating infinitesimal frame rotations on the GR sector of phase space, we find that
\begin{equation}
\underline{\mathcal{J}}[\tilde{\omega}]=\int_\Sigma dx\,
\omega_{ba}\underline{\mathcal{J}}^{[ab]}
\end{equation}
generates frame rotations on the Dirac sector
\begin{equation}
\bigl\{\boldsymbol{\psi},\underline{\mathcal{J}}[\tilde{\omega}]\bigr\}=
i\tfrac{1}{4}\hat{\omega}\boldsymbol{\psi}
=\Delta_{\tilde{\omega}}[\boldsymbol{\psi}],
%
%
\end{equation}
and the $\mathfrak{gl}(3,\mathbb{R})$ algebra~\eqref{eq:Jays} is reduced to that of $\mathfrak{so}(3)$
\begin{equation}\label{eq:Dirac Jays}
\bigl\{\underline{\mathcal{J}}[\tilde{N}],\underline{\mathcal{J}}[\tilde{M}]\bigr\}=
\int_\Sigma d^3x\, N_{ba}\Delta_{\tilde{M}}\bigl[\underline{\mathcal{J}}\bigr]^{[ab]}.
\end{equation}

The resulting E-D Hamiltonian is once again in standard form
\begin{equation}
H=\int_\Sigma d^3x\,\bigl(
N^\prime\underline{\mathcal{H}}^\prime
+{N^\prime}^a\underline{\mathcal{H}}^\prime_a
+N^\prime_{ba}\underline{\mathcal{J}}^{[ab]}
\bigr),
\end{equation}
where $\underline{\mathcal{H}}^\prime=\underline{\mathcal{H}}^{\prime\textsc{gr}}+\underline{\mathcal{H}}^{\textsc{d}}$ and $\underline{\mathcal{H}}^\prime_a=\underline{\mathcal{H}}^{\prime\textsc{gr}}_a+\underline{\mathcal{H}}^{\textsc{d}}_a$ (the GR contributions given by the primed constraints~($45$) of~\cite{Clayton:1997c} with $\gamma_{ab}$ replaced by $\delta_{ab}$), where 
\begin{equation}\label{eq:Hammom Dirac}
\underline{\mathcal{H}}^{\textsc{d}}=
-\tfrac{1}{2}iE\psi^\dagger\alpha^a
\overleftrightarrow{\nabla}_a\psi
+m\boldsymbol{\psi}^\dagger\beta\boldsymbol{\psi}
={\underline{T}^{\textsc{d}\perp}}_\perp
,\quad
\underline{\mathcal{H}}_a^{\textsc{d}}
=i\tfrac{1}{2}E\psi^\dagger\overleftrightarrow{\nabla}_a\psi
={\underline{T}^{\textsc{d}\perp}}_a
-E\nabla_b\bigl[{\mathcal{J}^{\textsc{d}}}^{[ab]}\bigr],
\end{equation}
arise from the Legendre transformation of~\eqref{eq:Dirac Lagrangian}.
Note that the equivalence of these with the E-D stress-energy tensor components~\eqref{eq:Dirac S-E} is achieved by making use of the Dirac equation~\eqref{eq:Dirac equation} to remove time derivatives of the spinor.
Also note that the presence of the derivative of the Dirac frame rotation generator in the momentum constraint is consistent with the relationship between the unprimed and primed constraints in~(45b) of~\cite{Clayton:1997c}.
From~\eqref{eq:Hammom Dirac} we find
\begin{equation}\label{eq:Dirac things}
\bigl\{\boldsymbol{\psi},
\underline{\mathcal{H}}^{\textsc{d}}[f]\bigr\}
=-fE^{1/2}\alpha^a\nabla_a[\psi]
-\tfrac{1}{2}\nabla_a[f]\alpha^a\boldsymbol{\psi}
-imf\beta\boldsymbol{\psi},\quad
\bigl\{\boldsymbol{\psi},
{\underline{\mathcal{H}}^\prime}^{\textsc{d}}_a[f^a]\bigr\}
=\pounds^\prime_{\vec{f}}[\boldsymbol{\psi}],
\end{equation}
where $\pounds^\prime$ acts as described in~\cite{Clayton:1997c}:
\begin{equation}
\pounds^\prime_{\vec{f}}[\boldsymbol{\psi}]
=E^{1/2}f^a\nabla_a[\psi]
+\Delta_{\tilde{\nabla f}}[E^{1/2}]\psi
=E^{1/2}f^a\nabla_a[\psi]+\tfrac{1}{2}\nabla_a[f]^a\boldsymbol{\psi}.
\end{equation}
Note that we have defined infinitesimal spatial diffeomorphisms to act on the frame as: $\{{E^i}_a,\underline{\mathcal{H}}^\prime_c[f^c]\}=-\nabla_a[f]^a{E^i}_b$; if one mixes in a spatial frame rotation so that only the symmetric part $\nabla_{(a}[f]_{b)}$ appears, then the resulting generator acting on $\psi$ reproduces the action that is used, for example, in~\cite{Henneaux:1980}.

From~\eqref{eq:Dirac things} one finds Hamilton's equations for $\psi$:
\begin{equation}
\bigl\{\boldsymbol{\psi},H\bigr\}
=E^{1/2}(N^a-N\alpha^a)\nabla_a[\psi]
+\tfrac{1}{2}(\nabla_a[N]^a-\alpha^a\nabla_a[N])\boldsymbol{\psi}
-imN\beta\boldsymbol{\psi}
+\tfrac{1}{4}i\hat{\omega}\boldsymbol{\psi},
\end{equation}
which are equivalent to~\eqref{eq:Dirac equation}.
Again making use of the definitions~\eqref{eq:definitions of things} we find (we have used $\{\sigma^{ab},\alpha^c\}=-2\epsilon^{abc}\gamma_5$)
\begin{subequations}\label{eq:Dirac uglies}
\begin{align}
{W^a}_b[\vec{f}]
&=-f^a\mathcal{H}^{\textsc{d}}_b
-\tfrac{1}{4}\nabla_c[f^c\psi^\dagger{\sigma_b}^a\psi]
-\tfrac{1}{4}\nabla_c[f^a\psi^\dagger{\sigma^c}_b\psi]
+\tfrac{1}{4}\nabla_c[f_b\psi^\dagger\sigma^{ac}\psi],\\
{Y^a}_b[f]
&=\tfrac{1}{2}if\psi^\dagger\alpha^a\overleftrightarrow{\nabla}_b\psi
+\tfrac{1}{8}\nabla_c[f\psi^\dagger\{\alpha^c,{\sigma_b}^a\}\psi],
\end{align}
\end{subequations}
and using~\eqref{eq:Dirac S-E} to find
\begin{equation}
\bar{T}^{\textsc{d}}_{ab}
=-\tfrac{1}{2}i\delta_{(ac}\psi^\dagger\alpha^c\overleftrightarrow{\nabla}_{b)}\psi
+\tfrac{1}{2}\delta_{(ac}K_{b)d}\psi^\dagger\sigma^{cd}\psi
+\tfrac{1}{2}\delta_{ab}m\psi^\dagger\beta\psi,
\end{equation}
we find that contributions to~\eqref{eq:pr second} and~\eqref{eq:pr third} are generated to properly reproduce the E-D field equations.
The constraint algebra combines~\eqref{eq:Dirac Jays} with~\eqref{eq:vacuum primed algebra}, replacing $\gamma_{ab}\rightarrow\delta_{ab}$ everywhere in the latter.
(Note that the derivation of the constraint algebra in~\cite{Clayton:1996b} could be generalized to the Einstein-Dirac system along the lines of~\cite{Lotze:1978}; note however the different parameterization of the spinor part of phase space and the resulting difference in the form of~\eqref{eq:Hammom Dirac}.)

\section{Discussion}

The intent of the preceding~\cite{Clayton:1997c} and present paper is to describe the Hamiltonian dynamics of general relativistic (vacuum and matter coupled) systems with respect to moving frames.
This extended system of dynamical spatial metric \textit{and} frame fields encompasses the following two natural limits:
If we choose the diffeomorphism constraints to act on the components of tensors with the usual Lie action and fix the spatial frame to be a coordinate frame, we recover the standard coordinate frame approach to canonical general relativity.
If we choose the spatial metric to be equal to the unit matrix and mix the diffeomorphism generators with the frame transformation generators to be compatible with this choice, we recover the orthonormal frame approach to canonical general relativity.
Clearly there are also a variety of intermediate choices available, however this enlarged arena provides a bridge over which one may pass from the standard coordinate frame approach to the Lorentz (or orthonormal) frame approach that is more natural for the description of the Einstein-Dirac system.

In this generalized structure, we find that matter fields na\"{\i}vely appear to be derivative-coupled since the momenta conjugate to the spatial frame fields are not (in general) independent of the matter fields.
This sort of derivative-coupling turns out to be benign since the field equations are equivalent to the coordinate frame approach to the model in question.
Indeed this feature is a necessary part of the frame approach, and will appear whenever the matter fields behave non-trivially under frame transformations.
In this way we understand that the derivative-coupling in the Einstein-Dirac system is a necessary feature, merely reflecting the need to work with local orthonormal frames in order to define the system.
We also end up with a simpler description of the Einstein-Dirac system than that given in Reference~\cite{Bao+Isenberg+Yasskin:1985} (note however their use of anti-commuting fermions), and a much more natural system than that found in Reference~\cite{Nelson+Teitelboim:1978} where primary importance is still given to the coordinate components of the metric tensor.
A more detailed examination of the Einstein-Dirac system in adjoint form (in the spirit of~\cite{Fischer+Marsden:1979}) is underway, and by adopting harmonic coordinate conditions and choosing the frame to fix the Local Lorentz gauge, we are able to prove local existence and uniqueness results.

\section*{Acknowledgements}

The author was supported by a postdoctoral fellowship from the Natural Sciences and Engineering Research Council of Canada, and thanks J. L\'{e}gar\'{e} for commenting on a preliminary form of this manuscript.

\providecommand{\bysame}{\leavevmode\hbox to3em{\hrulefill}\thinspace}

\end{document}